\documentclass{article}

\usepackage{mathrsfs}
\usepackage{svg}
\usepackage[ruled,linesnumbered]{algorithm2e}
\usepackage{verbatim}

\usepackage[cmex10]{amsmath}
\usepackage{amsthm,cite,graphicx,,booktabs,lipsum,color,bm,caption,subcaption,soul}
\usepackage{pifont,tikz,paralist,multirow,amssymb}
\usepackage{stmaryrd}
\usepackage{caption}
\usepackage{algorithmic}
\usepackage{mathtools}
\usepackage[labelformat=simple]{subcaption}

\usepackage{graphics}
\usepackage{xcolor}
\usepackage{siunitx}
\usepackage{tabularx}

\usepackage{PRIMEarxiv}

\usepackage[utf8]{inputenc} 
\usepackage[T1]{fontenc}    
\usepackage{hyperref}       
\usepackage{url}            
\usepackage{booktabs}       
\usepackage{amsfonts}       
\usepackage{nicefrac}       
\usepackage{microtype}      
\usepackage{lipsum}
\usepackage{fancyhdr}       
\usepackage{graphicx}       
\graphicspath{{media/}}     

\title{Deep Reinforcement Learning for Online Latency Aware Workload Offloading in Mobile Edge Computing
\thanks{\textit{\underline{Citation}}: 
\textbf{This paper has been accepted for the publication at the GLOBECOM' 22.}}
\thanks{This work was supported by the National Science Foundation under Award CNS-2148178.}
}

\author{
  Zeinab Akhavan, Mona Esmaeili, \\
  University of New Mexico \\
  Albuquerque\\
  \texttt{\{zakhavan, mesmaeili\}@unm.edu} \\
   \And
  Babak Badnava \\
  University of Kansas \\
  Lawrence\\
  \texttt{babak.badnava@ku.edu} \\
   \AND
   Mohammad Yousefi \\
   University of New Mexico \\
   Albuquerque \\
   \texttt{myousefi@unm.edu} \\
   \And
   Xiang Sun \\
   University of New Mexico \\
   Albuquerque \\
   \texttt{sunxiang@unm.edu} \\
   \And
   Michael Devetsikiotis \\
   University of New Mexico \\
   Albuquerque \\
   \texttt{mdevets@unm.edu} \\
   \And
   Payman Zarkesh-Ha \\
   University of New Mexico \\
   Albuquerque \\
   \texttt{pzarkesh@unm.edu} \\
}

\begin{document}
\maketitle

\begin{abstract}
Owing to the resource-constrained feature of Internet of Things (IoT) devices, offloading tasks from IoT devices to the nearby mobile edge computing (MEC) servers can not only save the energy of IoT devices but also reduce the response time of executing the tasks. However, offloading a task to the nearest MEC server may not be the optimal solution due to the limited computing resources of the MEC server. Thus, jointly optimizing the offloading decision and resource management is critical, but yet to be explored. Here, offloading decision refers to where to offload a task and resource management implies how much computing resource in an MEC server is allocated to a task. By considering the waiting time of a task in the communication and computing queues (which are ignored by most of the existing works) as well as tasks priorities, we propose the \ul{D}eep reinforcement l\ul{E}arning based offloading de\ul{C}ision and r\ul{E}source manageme\ul{NT} (DECENT) algorithm, which leverages the advantage actor critic method to optimize the offloading decision and computing resource allocation for each arriving task in real-time such that the cumulative weighted response time can be minimized. The performance of DECENT is demonstrated via different experiments.  

\end{abstract}


\section{Introduction}
The growing number of Internet of Things (IoT) devices, such as smart phones and smart watches, generate huge amount of data and tasks. Normally, some of these IoT devices are resource-constrained and do not have the capacity to process the tasks locally. The mobile cloud computing (MCC) technology has been proposed to allow these devices to offload their tasks to a remote data center. However, transmitting the tasks from IoT devices to a remote data center via the Internet is expensive, leading to high and uncontrollable latency \cite{Sun:2016:EdgeIoT}, thus unable to meet many IoT applications' requirements. For example, augmented reality requires the network delay to be less than 20 ms, which cannot be satisfied by MCC \cite{abrash2012latency}.     



To reduce the network latency, mobile edge computing (MEC) has been proposed to deploy many MEC servers at the network edge. Hence, instead of offloading tasks to a remote data center, IoT devices can offload their tasks to the nearby MEC servers offering low network delay, thus potentially reducing the response time. However, computing resources of MEC servers are limited, thus offloading a task to the nearest MEC server may not always be optimal because it may incur high computing latency of executing the task, although the network delay to offload the task is minimized. Many studies have designed methods to determine whether to offload tasks from the IoT devices under a dynamic environment \cite{Sun:2019:AAH,Zhang:2013:EMC}. This paper is built based on these methods by assuming the tasks have already been determined to be offloaded, but we are trying to solve the offloading decision problem, i.e., which MEC server should be selected to execute each of these task in a dynamic environment. Note that offloading decision and resource management are coupled together, meaning that whether an MEC server is suitable to execute a task depends on how much computing resource in the MEC server is allocated to the task, which is determined by the amount of remaining computing resource of the MEC server and the priority of the task. That is, if the task has low priority, i.e., low latency requirement, it is not necessary to assign all the remaining computing resource of the MEC server to the task.  

\begin{figure}[!htb]
	\centering	
	\includegraphics[width=.96\columnwidth]{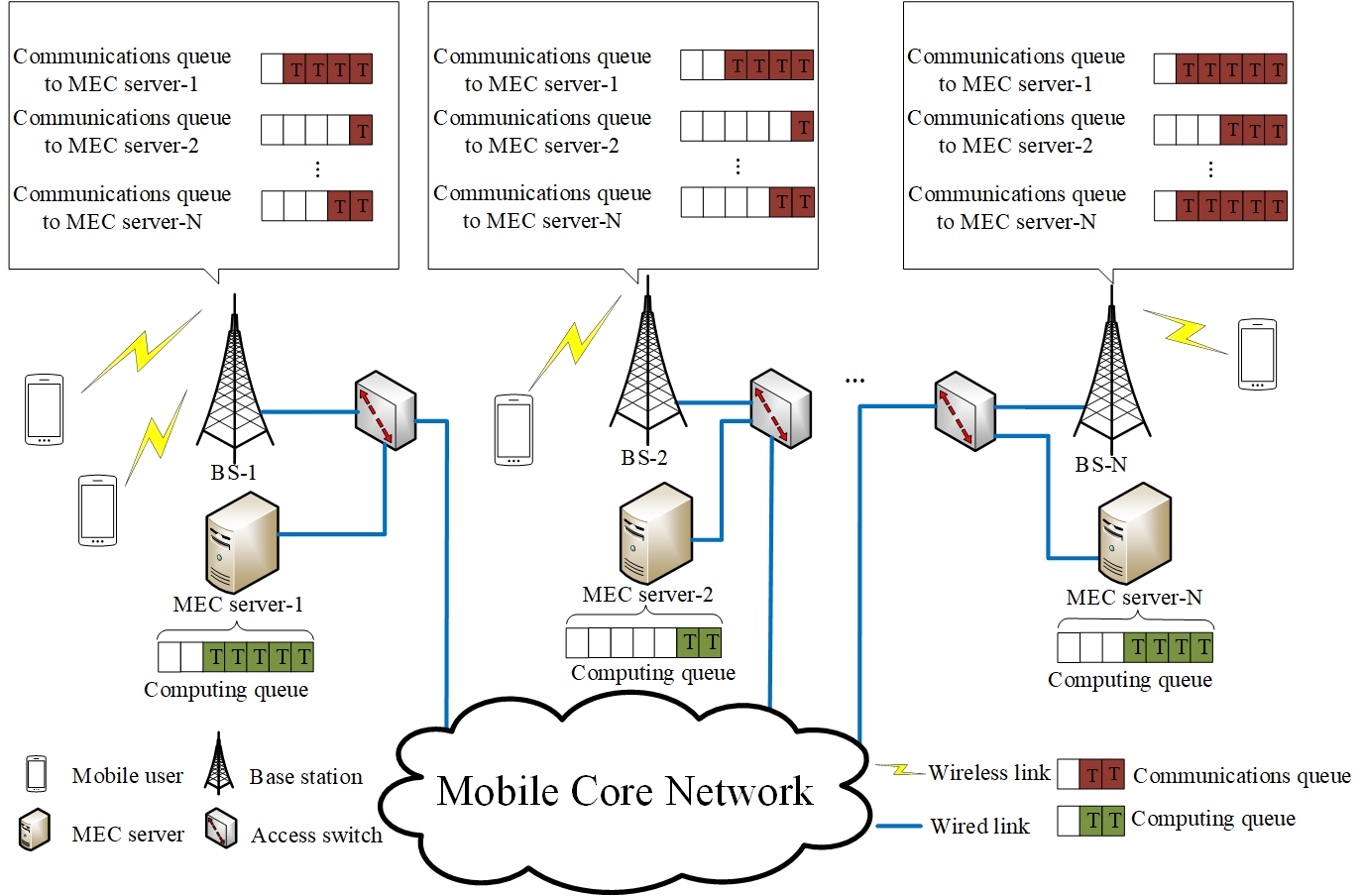} 
	\caption{The MEC architecture.}
	\label{fig:main_arch} 
\vspace{-10pt}
\end{figure}

To solve the joint offloading decision and resource management problem, machine learning and non-machine learning based solutions have been developed. Non-machine learning based solutions suggest a centralized controller to solve the optimization problem and determine the offloading and resource allocation of the incoming tasks at the BS \cite{sun2016primal,fan2018application,sun2017latency}. However, these solutions have the following drawbacks: 1) they only minimize the latency of the current IoT tasks by optimizing the offloading decision and resource allocation and do not consider the performance of the future IoT tasks, which may lead to the insufficient computing resources for the future IoT tasks at an MEC server, thus increasing their response time, and 2) they cannot make real-time decisions, i.e., the offloading decision and resource allocation cannot be made upon the arrival of a task. The existing machine learning based solutions employ deep reinforcement learning (DRL) to minimize the expected cumulative response time of all the tasks, which can resolve the second drawback of the non-machine learning based solutions \cite{Alfakih:2020:TOR,huang2020deep}. However, these solutions simplify the system by ignoring the waiting time of the tasks in the queues as well as the priorities of the tasks. Specifically, Fig.~\ref{fig:main_arch} shows the architecture of MEC, where each base station (BS) is attached to an MEC server via an access switch and maintains a number of communication queues, each of which buffers the arriving IoT tasks. Each MEC server executes the offloaded tasks and has a computing queue holding the tasks that are waiting for the computing resources to be released by the running tasks. The waiting time of an IoT task in the communication and computing queues would significantly affect the offloading decision and resource allocation, but yet to be considered in the existing solutions.

In this paper, we apply the advantage actor critic (A2C) method to solve the mentioned problem. Each BS observes the states of the system and determines the actions including the destination MEC server ID and the amount of the computing resources allocated to a task upon its arrival at the BS. The major contributions of the paper are listed as follows:
\begin{itemize}
    \item We formulate the joint offloading decision and resource management problem by considering different priorities of tasks and the waiting time of the tasks in the communication and computing queues. We model this optimization problem as Markov Decision Process (MDP). 
    \item We propose the \ul{D}eep reinforcement l\ul{E}arning based offloading de\ul{C}ision and r\ul{E}source manageme\ul{NT} (DECENT) algorithm to solve the problem based on A2C. 
    \item We demonstrate that DECENT outperforms the other two baseline approaches via extensive simulations.
\end{itemize}
 
The rest of this paper is organized as follows. Section~\ref{Sec:Related} presents related work. Section~\ref{Sec:Model} illustrates the related system model and presents the problem formulation of joint offloading decision and resource management. Section~\ref{Sec:Algorithm} provides the detail of the DECENT algorithm. Section~\ref{Sec:Simulation} discusses the simulation results, and Section~\ref{Sec:Conclusion} concludes the work.

\section{Related Work} \label{Sec:Related}
Many works focus on the strategy to determine a task should be offloaded to the nearby MEC server or executed locally such that the response time or the energy consumption of executing the task can be minimized \cite{liu2015adaptive,mukherjee2016power, Badnava-2021-spectrum}. For example, Elgazzar \emph{et al.} \cite{elgazzar2014cloud} proposed a decision model to evaluate whether offloading a task to the nearby MEC server improves its performance or not. The system operates by selecting a suitable resource provider to perform a task based on contextual information. Sun and Ansari \cite{sun2016primal} proposed a solution to place private virtual machines (VMs) with fixed computing resources for different mobile users to optimize the tradeoff between the migration gain and the migration cost. Assigning static computing resources to different VMs may lead to low resource utilization and increase the response time. By classifying the tasks into different IoT applications, Fan \emph{et al.} \cite{fan2018application} converted the task offloading problem into the application VM allocation problem. They proposed a method to dynamically adjust computing resources of different applications in each MEC server based on their workloads, thus reducing the computing delay of all tasks in the MEC server. Sun and Ansari \cite {sun2017latency}  proposed a Latency aware workload offloading algorithm to optimize the offloading decision such that the average response time of the tasks can be minimized. These two papers require a centralized server to obtain the tasks from different BSs and solve an optimization problem. Also, the waiting time of the tasks in the communications queue and the priorities of the tasks are not considered in these papers. 
Badnava \emph{et al.} \cite{Badnava-2021-spectrum} employed a Deep Q-Network to choose the best communication channel for task offloading to maximize the lifetime of a swarm of unmanned aerial vehicles (UAVs).
Jia \emph{et al.} \cite{jia2016cloudlet} aimed to balance the workload among different geo-distributed MEC servers such that the computing latency could be minimized. Yet, the network delay of transmitting the workload among the MEC servers is ignored. Alfakih \emph{et al.} \cite{Alfakih:2020:TOR}  applied the reinforcement learning approach to optimize the offloading decision and bandwidth allocation to minimize the system cost, which comprises energy consumption of a mobile device and computing delay of a task. However, the paper does not dynamically allocate computing resources to different tasks. Huang \emph{et al.} \cite{huang2020deep} proposed performance-aware resource allocation to efficiently assign computing and communication resources to users. The objective of the work is to maximize the long term performance of the system by using deep deterministic policy gradient (DDPG) to achieve the best resource allocation. 

\section{System Model} \label{Sec:Model}
Fig.  \ref{fig:main_arch} shows the MEC architecture, where each BS is attached to an MEC server and communicates with the IoT devices in the coverage area of the BS. Each BS maintains a number of communications queues, each of which holds the tasks waiting to be transmitted to the desired MEC server, and each MEC server maintains a computing queue that holds the tasks waiting for sufficient computing resources to be released on the MEC server. Each BS determines where to offload a task and how much computing resources is allocated to the task upon its arrival such that the average response time of a task is minimized. Let $\mathcal{I}$ and $\mathcal{K}$ be the sets of the tasks and MEC servers in the system, respectively. Let $i$ and $k$ be the indices of the tasks and MEC servers, respectively. Note that each BS is attached to an MEC server and we will use the same index to represent a BS and its attached MEC server. In general, the response time of offloading task $i$ to MEC sever $k$, denoted as $T_{ik}$, comprises the network delay $T^{net}_{ik}$ and computing delay $T^{comp}_{ik}$, i.e., $T_{ik}=T^{net}_{ik}+T^{comp}_{ik}$. The network delay $T^{net}_{ik}$ is the elapsed time from the arrival of task $i$ at the BS until its delivery to an MEC server $k$. Furthermore, the computing delay $T^{comp}_{ik}$ is the length of time from the arrival of task $i$ at MEC server $k$ to its completion. Without loss of generality, we do not consider the delay of sending the result of a task back to the IoT device. 

\textbf{Network Delay:}
The network delay of offloading task $i$ to MEC server $k$ comprises: 1) the transmission time $T^{trans}_{ik}$ of task $i$ from the BS to MEC server $k$ via the network, i.e., $T^{trans}_{ik}=\frac{{{l_i}}}{{{w_k}}}$, where $l_i$ is the size of task $i$ in bits and $w_k$ is the capacity of the path from the BS to MEC server $k$ in bps. 2) the E2E delay between the BS and MEC server $k$, denoted as $T^{e2e}_{ik}$, and 3) the waiting time of task $i$ in the communication queue from the BS to MEC server $k$, denoted as $T_{ik}^{wait\_net}$. Here, ${T^{e2e}_{ik}}$ can be measured and monitored by the network controller. For instance, software defined networking (SDN) can be applied to the mobile core network and thus the SDN controller can periodically monitor and record the E2E delay between any two endpoints \cite{YU:2015:SLM,Sun:2020:GCN}. Also, $T_{ik}^{wait\_net}$ is the time duration between the arrival of task $i$ at the BS and its transmission starting time to MEC server $k$, which approximately equals the sum of the transmission time of all the tasks in the communications queue, i.e.,
\begin{equation}
    T_{ik}^{wait\_net}=\sum_{i'\in \mathcal{I'}_{ik}}{\frac{l_{i'}}{w_k}},
\end{equation}
where $\mathcal{I'}_{ik}$ is the set of tasks in the communications queue for MEC server $k$ when task $i$ arrives at the BS. Hence, the network delay of offloading task $i$ to MEC server $k$ is 
\begin{align}
T_{ik}^{net}\!\!=\!T_{ik}^{trans}\!+\!T_{ik}^{e2e}\!+\!T_{ik}^{wait\_net}\!=\!\frac{l_i}{w_k}\!+\!T_{ik}^{e2e}\!+\!\!\!\sum_{i'\in \mathcal{I} '_{ik}}\!\!{\frac{l_{i'}}{w_k}}.
\label{Eq:1}
\end{align}

\textbf{Computing Delay:}
The computing delay of offloading task $i$ to MEC server $k$ comprises 1) the waiting time $T_{ik}^{wait\_comp}$ of task $i$ to be executed in the computing queue of MEC server $k$, and 2) the execution time $T_{ik}^{exe\_comp}$ of task $i$ at MEC server $k$. Normally, $T_{ik}^{exe\_comp}$ depends on the complexity of task $i$ (i.e., how many CPU cycles are required) and the amount of computing resources allocated to task $i$ in MEC server $k$, denoted as $r_{ik}$. The complexity of task $i$ is usually proportional to the size of task $i$ \cite{fan2021delay}, and thus we have $T_{ik}^{exe\_comp}=\frac{\mu l_i}{r_{ik}}$, where $\mu$ is the coefficient to map the size of a task in bits to the complexity of the task in CPU cycles. In addition, the waiting time $T_{ik}^{wait\_comp}$ depends on the number of tasks and their complexities in the computing queue when task $i$ arrives at MEC server $k$, i.e., $T_{ik}^{wait\_comp}=\sum_{i''\in \mathcal{I''}_{ik}}{\frac{\mu l_{i''}}{r_{i''k}}}$, where $\mathcal{I''}_{ik}$ is the set of tasks in the computing queue of MEC server $k$ at the arrival of task $i$. Hence, the computing delay of task $i$ to MEC server $k$ is
\begin{equation}
T_{ik}^{comp}=T_{ik}^{exe\_comp}+T_{ik}^{wait\_comp}=\frac{\mu l_i}{r_{ik}}+\sum_{i''\in \mathcal{I} ''_{ik}}{\frac{\mu l_{i''}}{r_{i''k}}}.
\end{equation}

\textbf{Problem Formulation:} 
We formulate the joint offloading decision and resource management problem as follows.
\begin{align}
\label{Eq:6}
		\textbf{P0:}\ \  &\mathop {\mathrm{arg}\min} \limits_{r_{ik},x_{ik}}\sum_{i\in \mathcal{I}}{\eta _i\left( \sum_{k\in \mathcal{K}}{x_{ik}\left( T_{ik}^{comp}+T_{ik}^{net} \right)} \right)},\\ 
		s.t.\ &\forall i \in \mathcal{I}, \sum\limits_{k \in \mathcal{K}} {x_{ik}}=1,  \label{const:c1}\\
		&\forall i \in \mathcal{I},\forall k \in \mathcal{K}, x_{ik}\in \left\{ {0,1} \right\}, \label{const:c2}\\
		&\forall i \in \mathcal{I},\forall k \in \mathcal{K}, r_{ik}=\left\{0,R_1,R_2,R_3,\cdots R^{\max} \right\} \label{const:c3},
\end{align}
where $x_{ik}$ is a binary variable indicating whether task $i$ is offloaded to MEC server $k$ ($x_{ik}=1$) or not ($x_{ik}=0$) and $\eta_i$ is the priority or weight of task $i$. A larger $\eta_i$ indicates the system prefers to reduce the response time of the task and vice versa. The objective of $\textbf{P0}$ is to minimize the overall weighted response time of all the tasks, where $\sum_{k\in \mathcal{K}}{x_{ik}\left( T_{ik}^{comp}+T_{ik}^{net} \right)}$ is the response time of task $i$. Constraint \eqref{const:c1} indicates any task can only be offloaded to a specific MEC server. Constraint \eqref{const:c2} means that $x_{ik}$ is a binary variable. Constraint \eqref{const:c3} defines the feasible values of $r_{ik}$, where $\left\{0,R_1,R_2,\cdots\right\}$ are the different computing resource blocks (e.g., the number of CPU cores) that can be allocated to task $i$ in MEC server $k$ and $R^{\max}$ is the maximum capacity of an MEC server.

$\textbf{P0}$ is nontrivial to be solved because 1) different tasks arrive at different time slots, and thus the BS cannot make the immediate decision for a task to minimize the delay if it is not aware of the future incoming tasks' information, and 2) $\textbf{P0}$ is an NP hard problem even if the BS can predict the information of all the tasks (i.e., their arrival time, complexities, and weights). Hence, we propose to apply a DRL method to find the sub-optimal solution of $\textbf{P0}$ in real-time.          

\section{Deep reinforcement learning based offloading decision and resource management} \label{Sec:Algorithm}
We apply MDP $\left( {\bm{\mathcal{S}},\bm{\mathcal{A}},\bm{\mathcal{F}},\bm{\mathcal{R}}} \right)$ to reformulate $\textbf{P0}$: 1) $\bm{\mathcal{S}}$ indicates the state space. A state at the arrival of task $i$, denoted as $\bm{s}_i \in \bm{\mathcal{S}}$, includes
\begin{itemize}
    \item Weight and data size of a new task, i.e., $\eta_i$ and $l_i$.
    \item Remaining computing resource of the MEC servers once new task $i$ arrives, i.e., $\bm{C}=\left\{ c_{k}\left| \forall k\in \bm{\mathcal{K}} \right. \right\}$, where $c_k$ is the remaining computing resource of MEC server $k$. Here, $c_{k}=c^{\max}-\sum_{j\in \bm{\mathcal{I}}_{ik}^{comp}}{r_{jk}}$, where $\bm{\mathcal{I}}_{ik}^{comp}$ is the set of tasks executed by MEC server $k$ once task $i$ arrives.
    \item Computing workload of MEC servers' computing queues once new task $i$ arrives, i.e., $\bm{D}=\left\{ d_{k}\left| \forall k\in \bm{\mathcal{K}} \right. \right\}$, where $d_{k}$ is the computing workload of MEC server $k$'s computing queue, i.e., $d_{k}=\sum_{j\in \bm{\mathcal{I}}_{ik}^{comp\_queue}}{z_{j}}$. $\bm{\mathcal{I}}_{ik}^{comp\_queue}$ is the set of tasks in MEC server $k$'s computing queue once new task $i$ arrives. 
    \item Waiting time of the communication queues once task $i$ arrives, i.e., $\bm{B}=\left\{ b_{k}\left| \forall k\in \bm{\mathcal{K}} \right. \right\}$, where $b_{k}$ is the waiting time of the communication queue for MEC server $k$, i.e., $b_{k}=\sum_{j\in \bm{\mathcal{I}}_{ik}^{comm\_queue}}{\frac{l_{j}}{w_k}}$. $\bm{\mathcal{I}}_{ik}^{comm\_queue}$ is the set of tasks in the communications queue once task $i$ arrives. 
\end{itemize}
2) $\bm{\mathcal{A}}$ is the set of actions for a BS to offload a task. The action set of task $i$, denoted as $\bm{a}_{i} \in \bm{\mathcal{A}}$, comprises $\bm{a}_i=\left\{r_{ik},x_{ik} \right\}$; 3) $\bm{\mathcal{F}}:\bm{\mathcal{S}} \times \bm{\mathcal{A}} \to \bm{\mathcal{S}}$ defines the state transition probability density function that maps the current states and actions into the next states; 4) $\bm{\mathcal{R}}: \bm{\mathcal{S}} \times \bm{\mathcal{A}} \to \mathbb{R}$ is the reward function. The reward function for task $i$ can be defined as the negative value of task $i$'s response time, i.e., $r_{i}=-\eta _i\left( \sum_{k\in \mathcal{K}}{x_{ik}\left( T_{ik}^{comp}+T_{ik}^{net} \right)} \right)$.

We then design the DECENT algorithm, which is based on Advantage Actor Critic (A2C) \cite{konda1999actor}, to solve the MDP problem. A2C is a DRL method combining policy-based and value-based reinforcement learning. In A2C, there are two neural networks, i.e., the actor and critic networks. The actor network provides the stochastic policy $\pi_\theta(\bm{a}_i|\bm{s}_i)$ to choose the actions $\bm{a}_i$ such that the expected cumulative reward, denoted as $J\left( \theta \right)$, is maximized. Here,
\begin{equation}
J\left( \theta \right) =\mathbb{E} \left[ \sum_{i'=i}^{\left| \mathcal{I} \right|}{\gamma^{i'}r_{i'}} \right],
\end{equation}
where $\theta$ is the parameter of the actor network, $\gamma \in [0,1]$ is the discount factor, and $\left| \mathcal{I} \right|$ is the total number of the tasks. According to \cite{A2C}, the gradient of $J\left( \theta \right)$ is: 
\begin{equation}
\nabla _{\theta}J(\theta )=\mathbb{E} [\nabla _{\theta}\log \pi _{\theta}(\bm{a}_i|\bm{s}_i)A(\bm{s}_i,\bm{a}_i)],
\end{equation}
where $A(\bm{s}_i,\bm{a}_i)$ is the Advantage function defined as
\begin{equation}
A(\bm{s}_i,\bm{a}_i)=r_{i}+\gamma V_{\upsilon}(\bm{s}_{i+1})-V_{\upsilon}(\bm{s}_i).
\end{equation}
Here, $V_{\upsilon}(\bm{s}_i)$ and $V_{\upsilon}(\bm{s}_{i+1})$ are the state-values with respect to task $i$ and $i+1$ estimated by the critic network and $\upsilon$ is the parameter of the critic network. Hence, the actor’s parameter $\theta$ is updated by the gradient descend, i.e.,
\begin{equation}
\label{eq:14}
\theta :=\theta -\beta _a\nabla _{\theta}J(\theta ),
\end{equation}
where $\beta _a$ is the learning rate of the actor network. 

The critic network in A2C is used to evaluate the actions taken by the actor network and provides the advantage value to the actor network to improve the policy. The objective of the critic network is to minimize the loss function $J(\upsilon)$, which is defined as the mean square error between the estimated state-value and the expected cumulative reward, i.e., 
\begin{equation}
J\left( \upsilon \right) =\left( r_{i}+\gamma V_{\upsilon}(\boldsymbol{s}_{i+1})-V_{\upsilon}(\boldsymbol{s}_i) \right) ^2
\end{equation}
Denote ${\nabla _{{\upsilon }}}J\left( {{\upsilon }} \right) $ as the gradient of $J\left( \upsilon \right)$ with respect to the parameter $\upsilon$. Then, $\upsilon$ is updated based on
\begin{equation}
\label{eq:16}
{\upsilon}:= {\upsilon}-\beta_c{\nabla _{{\upsilon }}}J\left( {{\upsilon }} \right) ,
\end{equation}
where $\beta _c$ is the learning rate of the critic network. 

The structures of the actor and critic networks are as follows. The actor network comprises an input layer taking in the input state $\bm{s}_i=\left\{ \eta_i,l_i,\bm{C},\bm{D},\bm{B} \right\}$, a hidden layer with 128 neurons and a relu activation function, and an output layer generating a probability distribution over actions with a softmax activation function. Likewise, critic network has an input layer taking in the state and action pair $\left<\bm{s}_i,\bm{a}_i \right>$, a hidden layer with 128 neurons and a relu activation function, and an output layer generating the state value.

Algorithm \ref{alg:DECENT} summarizes the DECENT algorithm, which is the process of training the actor and critic networks. Specifically, upon an arrival of task $i$ at the BS, the actor network applies the current policy to generate the action $\bm{a}_i=\left\{r_{ik},x_{ik} \right\}$ based on the current state $\bm{s}_i$. Note that we apply the $\epsilon$-greedy policy where it selects the random actions with the probability of 10\% and the greedy actions (that maximize the expected cumulative reward) with the probability of 90\%. Based on the actions $\bm{a}_i$, we calculate the corresponding reward $r_i$. This process is repeated until the actor and critic networks are converged. The well-trained actor and critic networks are used to determine the actions of incoming tasks in real-time.

\begin{figure}[!t]
\begin{algorithm}[H]
\label{alg:DECENT}
\SetAlgoLined
\caption{DECENT algorithm}
Initialize discount factor $\gamma$, learning rates $\beta_a$ and $\beta_c$, and exploration rate $\epsilon$.

\For {each training episode}{
    \For {each task arrival}{
         Obtain the current state $\bm{s}_i$;
         
        Input $\bm{s}_i$ to the actor network to calculate the actions $\bm{a}_i$ = [$x_{ik},r_{ik}]$ using $\epsilon$-greedy;
        
        Calculate reward $r_i$;

        Store transition $(s_i, a_i, r_i, s_{i+1})$ in the replay buffer;
   }
    
    Sample transitions from the replay buffer;
    
    Update the actor neural network based on Eq. \eqref{eq:14};
   
    Update the critic neural network based on Eq. \eqref{eq:16};
}
\end{algorithm}
\vspace{-10pt}
\end{figure} 

\section{Simulation Results} \label{Sec:Simulation}
In this section, we will conduct extensive simulations to validate the performance of the DECENT algorithm. Assume that there are one BS and 4 MEC servers located in different distances from the BS in $km$. The distances $d_k$ from the BS to 4 MEC servers are 0, 1, 2, and 3 km, respectively, where the distance is used to calculate the E2E delay $T^{e2e}_{ik}$ between the BS and an MEC server, i.e., $T_{ik}^{e2e} = \alpha \times d_k + \zeta$. Here, $\alpha$ and $\zeta$ are the coefficients, which are initially set to be 0.03 s/km and 0.03 s, respectively. The computing capacity of all the MEC servers are the same, i.e., $R^{max}=200$ Mcycles/s. The capacity of the links from the BS to the MEC servers are the same, i.e., $w_k=2\times 10^9$ bits/s. The arrival of tasks follows a Poisson distribution with the average arrival rate equaling to 50 tasks. The data size of an arrival task $l_i$ is randomly generated from a normal distribution, i.e., $l_i \sim N(3\times 10^7\ bits, 3\times 10^5)$. The computation intensity of a task $\mu=0.15$ CPU cycles/bit. In addition, the number of computing resource blocks that can be assigned to an incoming task $r_{ik}\in \{10,20,40,60,80,100,120,140,160,200\}$ Mcycles/s, and the weight of an arrival task $\eta_i$ is uniformly selected, i.e., $\eta_i \sim U\{10,20,50,100\}$, where a higher weight of a task implies the task has to be executed in a lower delay, and vice versa. Table \ref{tab:table2} shows other simulation parameters.            

The two baseline algorithms, i.e., nearest server and largest server, are used to compare the performance with DECENT. Here, the nearest server algorithm selects the closest MEC server (i.e., the lowest E2E delay) from the BS, and the largest server algorithm picks the MEC server with the largest remaining computing resource to offload a task.

\begin{table}[!htb]

\centering
\begin{tabular}{ll}
\toprule
\textbf{Parameter} & \textbf{Value}   \\ 
\midrule
Learning rate of the actor network $\beta_a$ & 0.0001 \\
Learning rate of the critic network $\beta_c$ & 0.0002\\ 
Exploration rate $\epsilon$  & 0.1      \\
Link capacity $w_k$     & $2\times 10^9$ bits/s \\
MEC server capacity $R^{max}$ & 200 Mcycle/s\\
\bottomrule
\end{tabular}%
\caption {\label{tab:table2} Simulation Parameters.} 
\vspace{-5pt}
\end{table}

\begin{figure}
\centering
\begin{minipage}{.5\textwidth}
  \centering
  \includegraphics[width=1.1\linewidth]{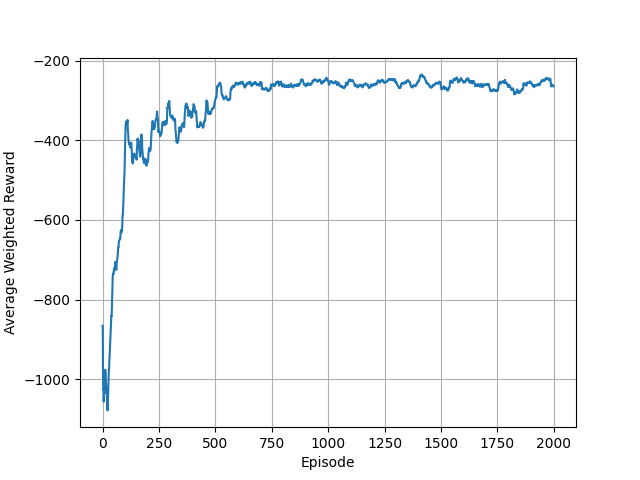}
  \captionof{figure}{Learning curve of DECENT.}
  \label{train}
\end{minipage}%
\begin{minipage}{.5\textwidth}
  \centering
  \includegraphics[width=1.1\linewidth]{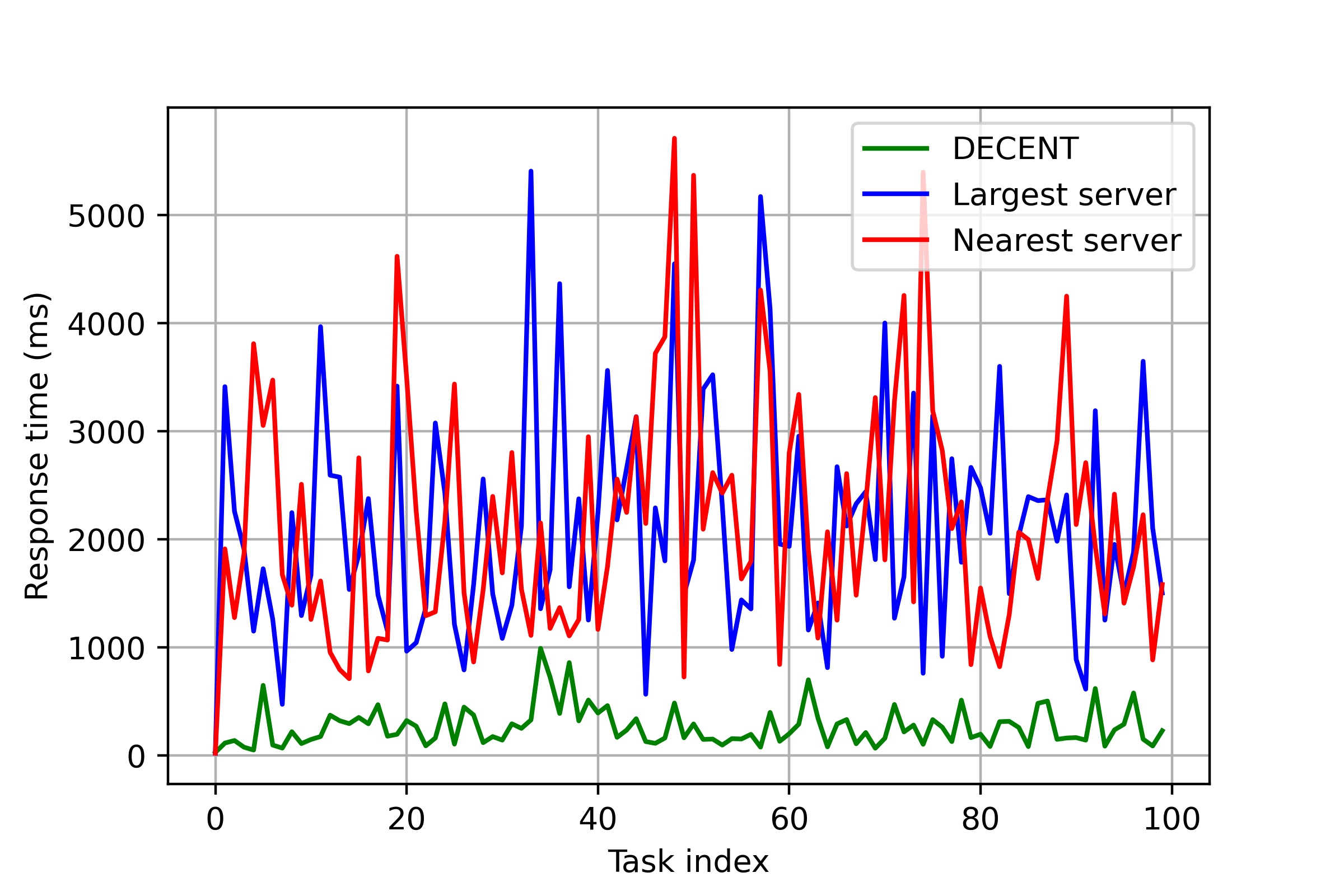}
  \captionof{figure}{Performance comparison.}
  \label{comparison}
\end{minipage}
\end{figure}

Fig. \ref{train} illustrates the learning curve for the DECENT algorithm, where DECENT can train the actor and critic networks to generate a better actions to maximize the average weighted reward in terms of minimizing the average weighted response time. The learning curves can quickly be converged around 500 episodes. Fig. \ref{comparison} compares the weighted response time of different tasks for different algorithms where the average task arrival rate $\lambda$ is 50 tasks/s. DECENT outperforms the baseline algorithms as the weighted response time of all the tasks can be lower than 1 second. The nearest MEC server and largest capacity MEC server algorithms, on the other hand, incurs a longer computing delay and communications delay, thus leading to a longer weighted response time. 



\begin{figure}
\centering
\begin{minipage}{.5\textwidth}
  \centering
  \includegraphics[width=1.1\linewidth]{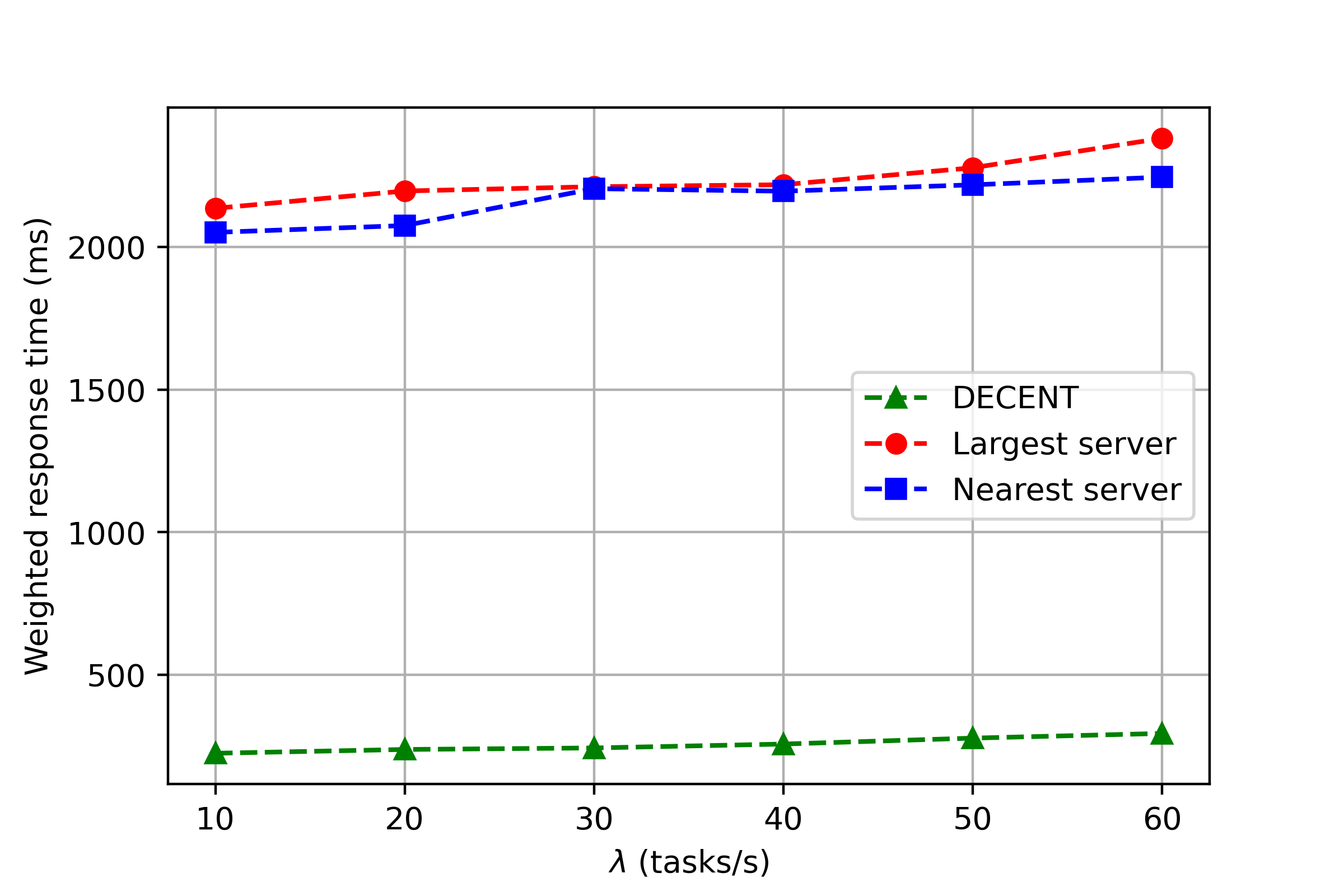}
  \captionof{figure}{Average weighted response time over average task arrival rate for different algorithms.}
  \label{lambda}
\end{minipage}%
\begin{minipage}{.5\textwidth}
  \centering
  \includegraphics[width=1.1\linewidth]{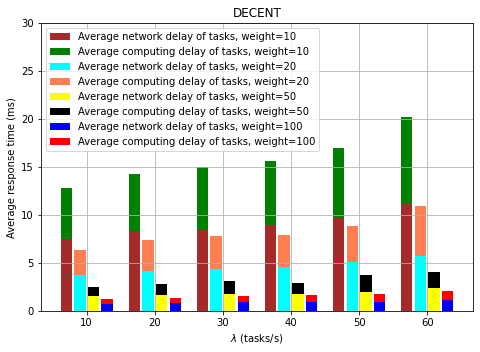}
  \captionof{figure}{Average response time w.r.t $\lambda$ and weights of the tasks.}
  \label{weight}
\end{minipage}
\end{figure}

We also examine the impact of the average task arrival rate $\lambda$ on the average weighted response time among 6,400 tasks. As shown in Fig. \ref{lambda}, as $\lambda$ increases, the average weighted response time incurred by DECENT slightly increases, while still maintaining a low level, i.e., $<500$ ms. The average weighted response time increment of the other two baseline algorithms over $\lambda$ is similar but much larger than DECENT, which demonstrates that DECENT achieves better offloading decision and resource management in both light and heavy workload scenarios. All the tasks can be clustered into four classes, and each class contains the tasks with the same weight. Fig. \ref{weight} shows the average network and computing delay of the tasks from different classes for DECENT. We can find that the tasks with higher weight incur lower execution time than the tasks with lower weight, which demonstrates that DECENT can adjust the offloading decision and resource allocation according to the weight of incoming tasks.



\begin{figure}
\centering
\begin{minipage}{.5\textwidth}
  \centering
  \includegraphics[width=1.1\linewidth]{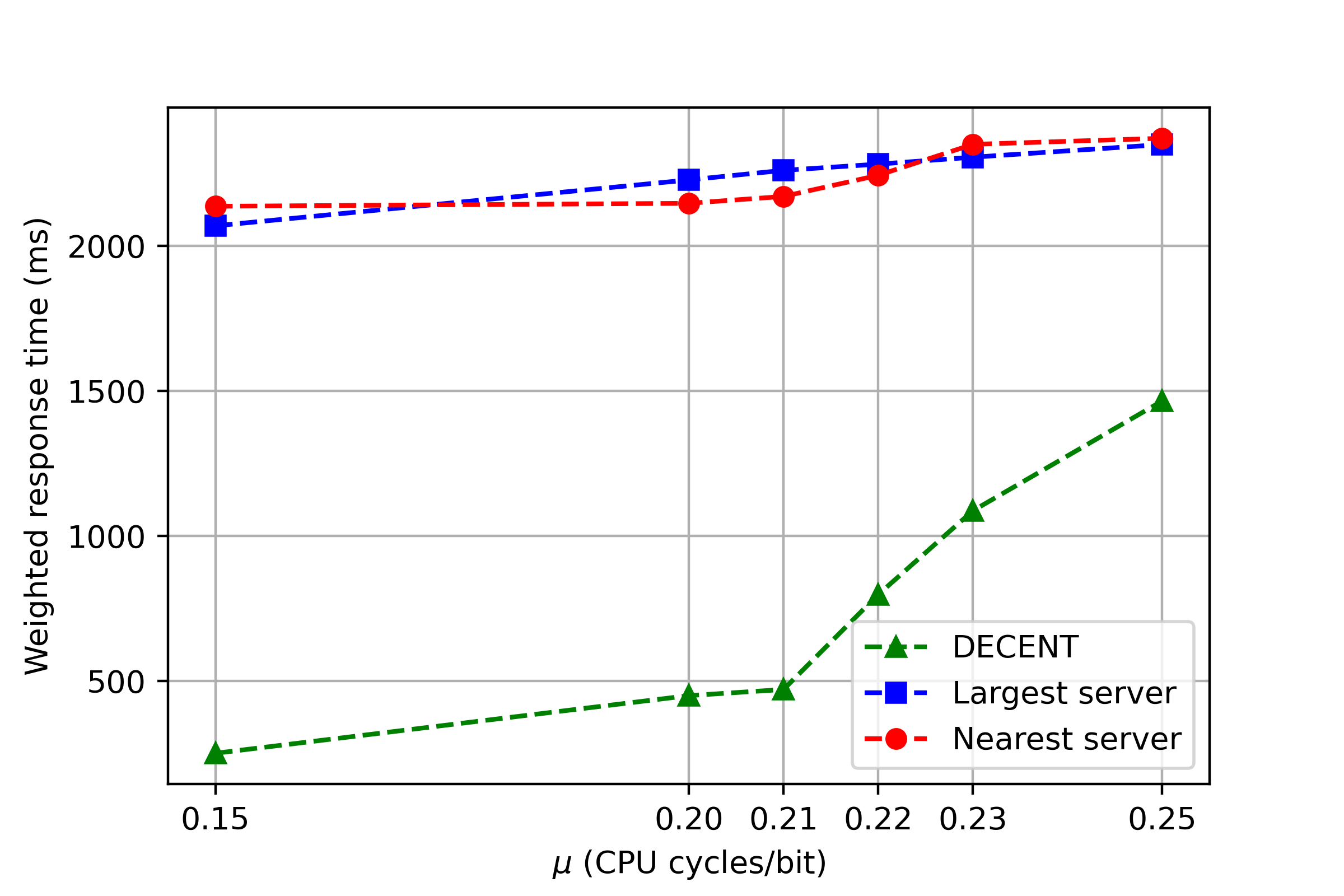}
  \captionof{figure}{Average weighted response time over $\mu$.}
  \label{mu}
\end{minipage}%
\begin{minipage}{.5\textwidth}
  \centering
  \includegraphics[width=1.1\linewidth]{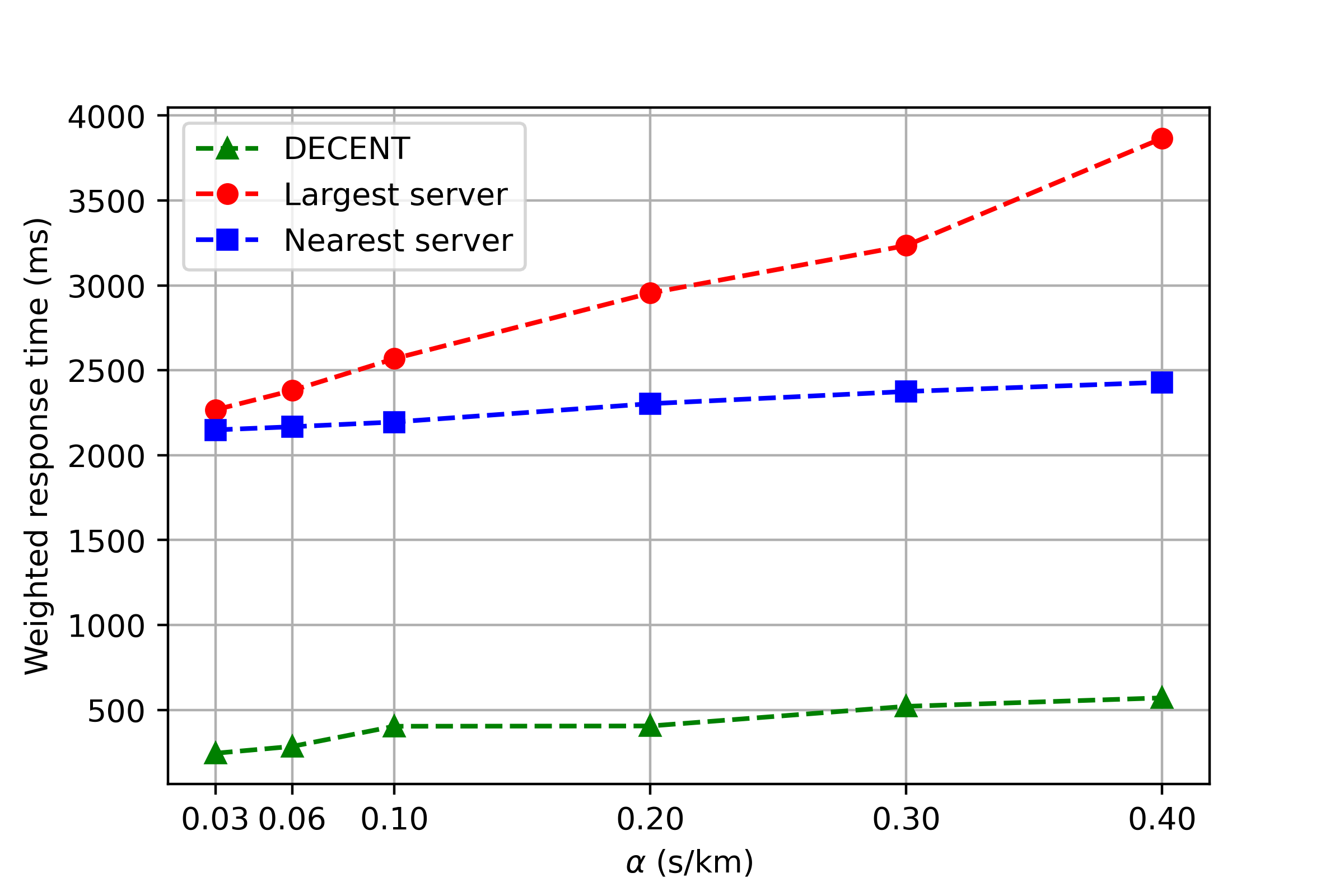}
  \captionof{figure}{Average weighted response time over $\alpha$.}
  \label{alpha}
\end{minipage}
\end{figure}

Figs. \ref{mu} and \ref{alpha} show the average weighted response time among 6,400 tasks by varying the computation intensity $\mu$ and E2E delay coefficient $\alpha$, respectively. Note that increasing $\mu$ and $\alpha$ would directly increase the E2E delay $T_{ik}^{e2e}$ and the execution time $T_{ik}^{exe\_comp}$, respectively. DECENT always incurs the lowest average weighted response time than the two baseline algorithms in different scenarios. It is interesting to see that the performance of the largest server algorithm in Fig. \ref{alpha} is significantly degraded as $\alpha$ increases. This is because increasing $\alpha$ increases $T_{ik}^{e2e}$, which may dominate the response time, and so offloading tasks to the nearby servers is preferred to reduce the weighted response time. Yet, the largest server algorithm does not consider the network delay, thus leading to significant performance degradation.

\section{Conclusion} \label{Sec:Conclusion}
By considering the waiting time of a task in the communication and computing queues as well as different priorities of the tasks, this paper proposed the DECENT algorithm allowing each BS to determine the offloading decision and computing resource allocation for each arrival task in real-time such that the cumulative weighted response time is minimized. As compared to the two baseline algorithms, DECENT has been demonstrated to have lower response time in different scenarios via extensive simulations. Also, DECENT is capable of adjusting the offloading decision and computing resource allocation based on the weights of the incoming tasks to minimize the cumulative weighted response time.

\bibliographystyle{unsrt}  
\bibliography{references}  
\nocite{*}

\end{document}